\documentclass[floatfix,groupedaddress,superscriptaddress,aps,pra, amsmath,amssymb,longbibliography, twocolumn,a4]{revtex4}
\usepackage{graphicx,float}% Include figure files
\usepackage{subfigure}
\usepackage{dcolumn}% Align table columns on decimmal point
\usepackage{bm}% bold math
\usepackage{mathrsfs}
\usepackage{txfonts}
\usepackage{CJK}
\usepackage{epsfig}
\usepackage{lipsum}
\usepackage{color}
\usepackage{placeins}
\usepackage[colorlinks]{hyperref}
\usepackage{xcolor}

\makeatletter
\emergencystretch=\maxdimen
\newcommand{\Rmnum}[1]{\expandafter\@slowromancap\romannumeral #1@}
\makeatother

\hypersetup{ plainpages=true,
    breaklinks=true,    hypertexnames=false,    pageanchor=true,
    colorlinks=true,
    linkcolor={blue},
    citecolor={blue},
    urlcolor={blue},
    anchorcolor={black}
}
\begin{document}
%\definecolor{backgroundcolor}{RGB}{199, 238, 206} 
%\pagecolor{backgroundcolor}

\title{Quantum synchronization via Active-Passive-Decomposition configuration: An open quantum system study}

\author{Nan Yang}\email{nyang.hust@gmail.com}
\affiliation{Center for Quantum Science and Engineering, and Department of Physics, Stevens Institute of Technology, Hoboken, New Jersey 07030, USA}

\author{Ting Yu}
\affiliation{Center for Quantum Science and Engineering, and Department of Physics, Stevens Institute of Technology, Hoboken, New Jersey 07030, USA}
\date{\today}% It is always \today, today,
%             %  but any date may be explicitly specified

\begin{abstract} 
In this paper, we study the synchronization of dissipative quantum harmonic oscillators in the framework of quantum open system via the Active-Passive Decomposition (APD) configuration. We show that
two or more quantum systems may be synchronized when the quantum systems of interest are embedded in dissipative environments and influenced by a common classical system. Such a classical 
system is typically termed as a controller, which (1) can drive 
quantum systems to cross different regimes (e.g., from periodic to chaotic motions) and (2) constructs the so-called Active-Passive Decomposition configuration such that all the quantum objects 
under consideration may be synchronized. The main findings of this paper is that we demonstrate that the complete synchronizations measured by the standard quantum deviation may be achieved for
both stable regimes (quantum limit circles) and unstable regimes (quantum chaotic motions). 
As an example, we numerically show in an optomechanical setup that the complete synchronization can be realized in quantum mechanical resonators.

% \textcolor{red}{The criteria for quantum synchronization of two quantum systems are also discussed in this paper, we utilize standard deviations of the position and momentum of each quantum system, to measure %quantum synchronization.} Physically, quantum synchronization in this work is understood as synchronous changes of quantum uncertainties. 
\end{abstract}

%\pacs{}
%\keywords{Suggested keywords}%Use showkeys class option if keyword
                              %display desired
\maketitle

%\tableofcontents

\section{Introduction}

The history of the synchronization can be traced back to the Dutch scientist, C.~Huygens, who firstly recorded the synchronization behaviors of two suspended pendulums~\cite{Huygens}. He also accurately understood the physics behind the phenomena: the oscillators may adjust their rhythms to reach a consistent state due to weak interactions. 
Since then, the synchronization problems have been recognized as a universal phenomenon in nature ranging from bursting neurons, fireflies, and chemical reactions, to  human activities, seasonal migrations, and solar systems, and the references therein~\cite{review1,review2}. In the domain of classical physics, for example, the synchronization of periodic oscillators may be studies based on the Adler equation~\cite{Adler1946} and the Kuramoto model~\cite{Kuramoto1975}. In addition, the chaotic synchronization has also been extensively studied by many different methods~\cite{Pecora1990, APD1,APD2,PS1,PS2,GS1,GS2,LS1}. Among them, the so-called Active-Passive Decomposition (APD) configuration~\cite{APD1,APD2} provides a general method for the complete synchronization of chaotic systems. 

The synchronization problems in quantum systems are also of great interests due to their practical applications and close connections with the fundamental aspects of quantum physics such as {the} transition from quantum to classical domains. The related works have been reported in various quantum systems including atomic systems~\cite{Xu2014,Karpat2020,Siwiak-Jaszek2020,Jaseem2020}, qubits~\cite{Zhirov2008,Zhirov2009,Giorgi2016,Cattaneo2021}, cold irons~\cite{Hush2015}, spin  models~\cite{Orth2010,Bellomo2017,Roulet2018,Giorgi2013,Karpat2019,Tindall2020,Goychuk2006}, lattices and dimer~\cite{Cabot2019,Michailidis2020,Siwiak-Jaszek2019}, Van der Pol  oscillators~\cite{Lee2013,Lee2014,Walter2015,Sonar2018,Eneriz2019,Mok2020,Kato2021}, 
quantum harmonic  oscillators~\cite{Giorgi2012,Manzano2013,Benedetti2016,Davis2016,Walter2014,Makino2016,Lorch2016,Lorch2017,Nigg2018,Qiao2018,Wachtler2020,Koppenhofer2020}, and optomechanical systems\cite{Amitai2017,Liao2019,Heinrich2011,Ludwig2013,Weiss2016,Li2016}. 
 Quantum synchronization in non-Markovian environments were studied in Ref.~\cite{Karpat2021}.  It has been experimentally demonstrated that high frequency resonators may potentially enable  the quantum synchronization~\cite{experiments1,experiments2,experiments3,experiments4,experiments5,experiments6,experiments7}. More applications of quantum synchronization in quantum network and key-distribution may be found in ~\cite{Kwasigroch2017,Pljonkin2017}. 
Despite remarkable progress, many interesting problems involving quantum open systems  still need to be carefully studied. Particularly,  how to synchronize quantum systems in unstable regimes is
a long-standing problem.

%}
%So far, the studies of quantum synchronization are restricted to stable quantum systems such as limit circles, how to synchronize quantum systems in unstable regimes is yet to be solved. 

\begin{figure}[t]
	\centering
	\includegraphics[width=3.4 in]{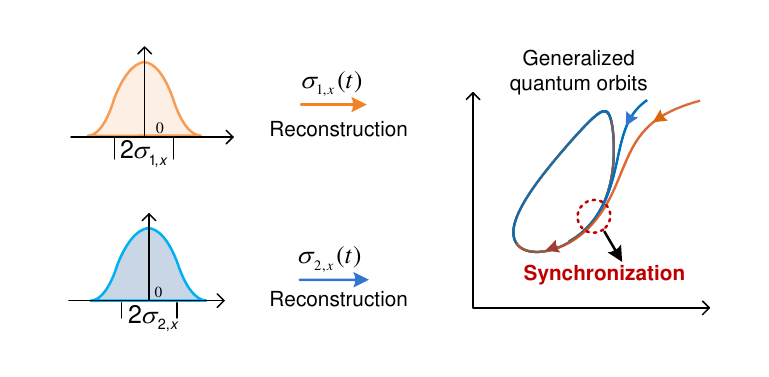} 
	\caption{(color online) {{Criterion} for quantum synchronization of two systems $1$ and $2$ is characterized by the identity of their standard deviations ($\sigma_{1,x}, \sigma_{1,p}$) and ($\sigma_{2,x}, \sigma_{2,p}$).}  
	}
	\label{Fig1}
\end{figure}

The purpose of this paper is {studying} the synchronization of two quantum harmonic oscillators in the quantum open system framework. We demonstrate the quantum synchronization may be achieved for the
dissipative quantum systems in not only stable regimes, but also the more complex unstable regimes where the chaotic dynamics is involved.

For our quantum harmonic oscillator model, it is convenient to use the standard deviations $\sigma_{x}$,  $\sigma_{p}$ of the position and momentum operators $\hat{x}$, $\hat{p}$ to measure the degree of quantum synchronization. As shown in this paper,  the given quantum synchronization criterion allows us to  analytically demonstrate the synchronization between two quantum harmonic oscillators in two separate  dissipative environments. Our approach have used the Active-Passive Decomposition (APD) configuration~\cite{APD1,APD2} to show that the participating classical system can enable the quantum synchronization through constructing an APD configuration and drives the quantum systems into a desired type of motion (e.g., from period to chaos). For the continuous-variable  systems  considered in this paper, we can reconstruct classical trajectories from the quantum  systems~\cite{Takens,Nan2020}.  Finally, we also numerically show in an experimentally accessible optomechanical setup~\cite{Aspelmeyer2014,Sciamanna2016,Monifi2016,Bakemeier2015,Buters2015,Carmon2007,Larson2011,Lee2009,Lv2015,MWang2016,Ma2014,Marino2013,
 	Navarro-Urrios2017,Piazza2015,Sun2014,Suzuki2015,Walter2016,WangGuanglei2014, Liao2013,
 	WuJiagui2017,YangNan2015,ZhangK2010,ZhangJing2021,LiuYulong2017,WangXin2019,QinWei2021} that the desirable synchronization can be accomplished 
.

This paper is organized as follows.  In Sec.~II, we introduce the criterion of quantum synchronization for our continuous-variable systems. In Sec.~III, we discuss the realization of quantum synchronizations
in several interesting scenarios. Then, in Sec.~IV, we apply our quantum synchronization approach to an optomechanical setup; our numerical simulations show that the quantum synchronization is achieved in two quantum mechanical resonators in both periodic and chaotic regimes. Finally, we summarize and conclude this paper in Sec.~V.

\section{Criteria for quantum synchronization} 

In classical dynamics, the synchronization of two oscillators can be measured by the relative relationships between their trajectories, e.g., 
complete synchronization refers to approaching identical trajectories. 
Quantum systems by default would not allow us to use concepts like "classical trajectories" directly~\cite{Mari2013}. 
%This prevents quantum synchronization being defined in the framework of the existing synchronization theory. 
In quantum domains, various definitions for synchronization have been proposed, such as the semiclassical trajectory approach~\cite{Mari2013},  the relative phase distribution~\cite{Hush2015} and Wigner function approach~\cite{Makino2016}. Also, it should be noted that a definition based on observables has been provided recently for the synchronization of quantum limit circles~\cite{Buca2022}.

For our {discussions}, the criteria to be used to measure the quantum synchronization for continuous variables are based on the standard deviations [$\sigma_{x}(t)$, $\sigma_{p}(t)$] of the position and momentum operators ($\hat{x}$, $\hat{p}$) 
of the harmonic oscillator systems: $\sigma_{x}=\sqrt{\langle \hat{x}^2 \rangle - \langle \hat{x} \rangle^2}$ and $\sigma_{p}=\sqrt{\langle \hat{p}^2 \rangle - \langle \hat{p} \rangle^2}$. Of course if the quantum system
under consideration has other degrees of freedom such as spins or discrete systems, our definition of the synchronization must be modified accordingly~\cite{Xu2014,Karpat2020,Siwiak-Jaszek2020,Jaseem2020,Zhirov2008,Zhirov2009,Giorgi2016,Cattaneo2021,Hush2015,Orth2010,Bellomo2017,Roulet2018,Giorgi2013,Karpat2019,Tindall2020,Goychuk2006}. In this way, the synchronization of two quantum systems may be measured by the relative relationships between their standard deviations [$\sigma_{1,x}(t)$, $\sigma_{1,p}(t)$] and [$\sigma_{2,x}(t)$, $\sigma_{2,p}(t)$] of the quantum position and momentum ($\hat{x}$, $\hat{p}$).

To introduce the detailed definition for the quantum synchronization, we first review the definition of complete synchronization in classical systems~\cite{review1,Pecora1990}.

\emph{Complete synchronization} in classical contexts refers to the identity of the trajectories of two dynamical systems. 
Consider two autonomous dynamical systems 
{$\dot{\textbf{y}}_1=\textbf{f}(\textbf{y}_1)$ and $\dot{\textbf{y}}_2=\textbf{f}(\textbf{y}_2)$},
where $\textbf{y}_1$ and $\textbf{y}_2$ are $N$-dimensional
variables governed by the function $\textbf{f}: R^N \rightarrow
R^N$. Here, $\textbf{y}_1$ and $\textbf{y}_2$ are called complete synchronization if and only if their difference
$\lim_{t\rightarrow\infty}\|\textbf{y}_1(t)-\textbf{y}_2(t)\|=0$. 

In the similar spirit, here we show that the above definition can be directly extended to quantum systems when the standard deviations are used. 

(\textbf{Complete synchronization}) Two 
quantum systems $1$ and $2$ are called complete synchronization if their standard deviations [$\sigma_{1,x}(t)$, $\sigma_{1,p}(t)$] and [$\sigma_{2,x}(t)$, $\sigma_{2,p}(t)$] satisfy the conditions: 
\begin{equation}\label{condition1}
\lim_{t \rightarrow \infty}\|\sigma_{1,x}(t)-\sigma_{2,x}(t)\|=0,
\ \ \  
\lim_{t \rightarrow \infty}\|\sigma_{1,p}(t)-\sigma_{2,p}(t)\|=0, 
\end{equation} 
where [$\sigma_{1,x}(t)$, $\sigma_{1,p}(t)$] and [$\sigma_{2,x}(t)$, $\sigma_{2,p}(t)$] represent the standard deviations of the two quantum systems $1$ and $2$.

In the next section, we will use these criteria to discuss some physically interesting examples that can display the realization of quantum synchronizations.

\section{Synchronization of two quantum harmonic oscillators based on Active-Passive Decomposition configuration}
\begin{figure}[t]
	\centering
	\includegraphics[width=3 in]{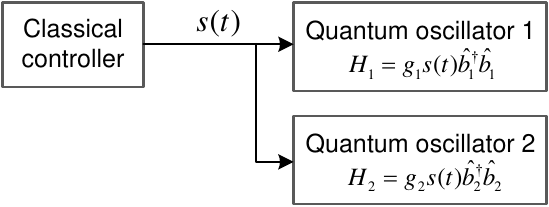}
	\caption{(color online) The schematic diagram for the synchronization of two quantum harmonic oscillators $\hat{b}_{1}$ and $\hat{b}_{2}$.}
	\label{Fig2}
\end{figure}

%With the criteria of quantum synchronization clarified, we move to the question of how to achieve quantum synchronization. 

To begin with, let us notice that, for classical dynamical systems, the so-called complete synchronization may be accomplished through many different ways such as mutual interactions, common driving forces, and 
feedback control mechanisms, etc. Interestingly, the synchronization has been extended to some important nonlinear systems where the chaotic synchronization may be observed. Among several useful methods
in realizing chaotic  synchronization, we find the Active-Passive Decomposition ({\rm APD}) configuration particularly convenient for our discussions. 
%In this work, we propose an approach for the synchronization of quantum harmonic oscillators in the framework of the Active-Passive Decomposition ({\rm APD}) configuration.}  

To put our discussions into perspective, let us first briefly review the basic idea of Active-Passive Decomposition (APD) configuration. 
In the {\rm APD} model, two 
chaotic subsystems to be synchronized can be written as 
non-autonomous forms: 
\begin{equation}
\dot{\textbf{z}}_1=\textbf{f}[\textbf{z}_1,\textbf{s}(t)], \qquad
\dot{\textbf{z}}_2=\textbf{f}[\textbf{z}_2,\textbf{s}(t)],
\end{equation}
where the dynamics of both  $\textbf{z}_1$ and $\textbf{z}_2$
are ruled by the function $\textbf{f}$, and $\textbf{s}(t)$ is the
common external driving governed by the autonomous function
$\dot{\textbf{s}}(t)=\textbf{h}[\textbf{s}(t)]$.
{
Here, the synchronization differences $\textbf{e}$ is defined as $\textbf{e}=\textbf{z}_1-\textbf{z}_2$, and its dynamics  is ruled by}
	\begin{equation}
	\dot{\textbf{e}}=\textbf{f}[\textbf{z}_1,\textbf{s}(t)] - \textbf{f}[\textbf{z}_1 -  \textbf{e},\textbf{s}(t)].
	\label{error_equation}
	\end{equation}
	{
The APD model constructs a configuration that Eq.~(\ref{error_equation}) is asymptotically stable at the zero point $\textbf{e}=0$. Thus, the synchronization differences $\textbf{e}$ goes to zero as the time increases and  
complete synchronization occurs for two chaotic dynamical systems $\textbf{z}_1$ and $\textbf{z}_2$.} 

As to be shown below,  the Active-Passive Decomposition (APD) configuration~\cite{APD1,APD2} may also be applied to the synchronization of quantum systems. 
As shown in Fig.~\ref{Fig2}, 
our quantum  model consists of two separate quantum harmonic oscillators ($\hat{b}_{1}$ and $\hat{b}_{2}$) {in different heat baths} and a classical controller that produces a common force $s(t)$ acting on $\hat{b}_{1}$ and $\hat{b}_{2}$ simultaneously. Here, the effective Hamiltonian of the two harmonic oscillators is given by 
\begin{equation}\label{H}
H=\hbar[\Omega_{1} + g_1s(t)] \hat{b}^\dag_{1} \hat{b}_{1} + \hbar[\Omega_{2} + g_2s(t)] \hat{b}^\dag_{2} \hat{b}_{2},
\end{equation}
where $\Omega_{j}$ is the resonant frequency of the $j-th$ quantum harmonic oscillator $\hat{b}_{j}$. The term $g_js(t)$ refers to the frequency shift of $\hat{b}_{j}$ brought by the classical input $s(t)$, and  $g_j$ is the coupling strength. In this setting, the quantum dynamics are modulated by the classical controller, which can {be driven} from periodic to chaotic regimes by adjusting the classical input $s(t)$. In what follows, we first derive the equations of motions for the second order terms of the quantum harmonic oscillators ($\hat{b}_{1}$ and $\hat{b}_{2}$), which determine the dynamics of {the corresponding} standard deviations [$\sigma_{1,x}$, $\sigma_{1,p}$] and [$\sigma_{2,x}$, $\sigma_{2,p}$]. Then, we will demonstrate that the quantum synchronization of two harmonic oscillators is achievable and stable in the  {\rm APD} configuration. 

\subsection{The equations of motions for the second order terms of the quantum harmonic oscillators $\hat{b}_1$ and $\hat{b}_2$}

For convenience, we use the shifted quantum harmonic oscillators $\hat{b}_{1}$ and $\hat{b}_{2}$ 
\begin{equation}\label{a_2b}
\hat{b}_1=\beta_1 + \hat{\tilde{b}}_1, \ \ \ \hat{b}_2=\beta_2 + \hat{\tilde{b}}_2,
\end{equation}
where $\beta_j = \langle \hat{b}_j \rangle$ refers to the classical mean value and $\hat{\tilde{b}}_j$ is the quantum fluctuation term of the mechanical mode $\hat{b}_j$  ($\langle \hat{\tilde{b}}_j \rangle = 0$). 
Then, the Hamiltonian in terms of $\hat{\tilde{b}}_j$ $(j=1,2)$ is obtained by substituting Eq.~(\ref{a_2b}) into Eq.~(\ref{H})
\begin{equation}\label{H_fluctuation}
H=\hbar\Omega^{\prime}_{1}(t) \hat{\tilde{b}}_{1}^\dag \hat{\tilde{b}}_{1} + \hbar\Omega^{\prime}_{2}(t) \hat{\tilde{b}}_{2}^\dag \hat{\tilde{b}}_{2},
\end{equation}
where $\Omega_j^\prime(t)=\Omega_{j} + g_js(t)$ is the modified resonant frequency of the $j-th$ quantum harmonic oscillator $\hat{b}_j$ for $j=1,2$. 
With the Hamiltonian given by Eq.~(\ref{H_fluctuation}), we now have the system master equation,
\begin{equation}\label{meq}
\begin{split}
\dot{\rho} =&i\left[\rho,\hbar\Omega_{1}^{\prime}(t) \hat{\tilde{b}}_{1}^\dag \hat{\tilde{b}}_{1} + \hbar\Omega_{2}^{\prime}(t) \hat{\tilde{b}}_{2}^\dag \hat{\tilde{b}}_{2}\right] \\
&+\sum_{j=1,2} \left[\Gamma_j [n_{j,{\rm th}}(t)+1]\left(2\hat{\tilde{b}}_j\rho\hat{\tilde{b}}_j^{\dag}-\hat{\tilde{b}}_j^{\dag}\hat{\tilde{b}}_j\rho - \rho\hat{\tilde{b}}_j^{\dag}\hat{\tilde{b}}_j\right) \right]\\
&+ \sum_{j=1,2} \left[\Gamma_j n_{j,{\rm th}}(t)\left(2\hat{\tilde{b}}_j^{\dag}\rho\hat{\tilde{b}}_j-\hat{\tilde{b}}_j\hat{\tilde{b}}_j^{\dag}\rho - \rho\hat{\tilde{b}}_j\hat{\tilde{b}}_j^{\dag}\right) \right], 
\end{split}
\end{equation}
where $\Gamma_j$ is the damping rate of the $j-th$ quantum harmonic oscillator $b_j$, and its mean thermal photon (phonon) excitation number $n_{j,{\rm th}}(t)=\exp{[{\hbar\Omega_j^\prime}(t)/{\kappa_{\rm B}T_j}-1]^{-1}}$ is determined by the temperature $T_j$ and the effective resonant frequency $\Omega_j^\prime(t)$. Here, $\kappa_{\rm B}$ is the Boltzmann constant. From the master equation, we can then obtain the equations of motions for $\langle \hat{\tilde{b}}_1^\dag \hat{\tilde{b}}_1 \rangle$, $\langle \hat{\tilde{b}}_2^\dag \hat{\tilde{b}}_2 \rangle$, $\langle \hat{\tilde{b}}_1^\dag \hat{\tilde{b}}_2 \rangle$, $\langle \hat{\tilde{b}}_1 \hat{\tilde{b}}_2 \rangle$, $\langle \hat{\tilde{b}}_1^2 \rangle$, and $\langle \hat{\tilde{b}}_2^2 \rangle$ by applying ${\langle\dot{\hat{o}}\rangle}={\rm Tr}(\rho\hat{o})$ for the operator $\hat{o}$,
\begin{subequations}\label{2nd_eq1}
	\begin{align}
		\frac{d\langle \hat{\tilde{b}}_1^\dag \hat{\tilde{b}}_1 \rangle}{dt} =& -\Gamma_1 \langle \hat{\tilde{b}}_1^\dag \hat{\tilde{b}}_1 \rangle+\Gamma_1 n_{j,\rm th}[\Omega_1^\prime(t)], \\
		\frac{d\langle \hat{\tilde{b}}_2^\dag \hat{\tilde{b}}_2 \rangle}{dt} =& -\Gamma_2 \langle \hat{\tilde{b}}_2^\dag \hat{\tilde{b}}_2 \rangle+\Gamma_2 n_{j,\rm th}[\Omega_2^\prime(t)], \\
		\frac{d\langle \hat{\tilde{b}}_1^\dag \hat{\tilde{b}}_2 \rangle}{dt} =& -i [-\Omega_1^{\prime}(t)+\Omega_2^{\prime}(t)] \langle \hat{\tilde{b}}_1^\dag \hat{\tilde{b}}_2 \rangle-\frac{\Gamma_1+\Gamma_2}{2} \langle \hat{\tilde{b}}_1^\dag \hat{\tilde{b}}_2 \rangle, \\
		\frac{d\langle \hat{\tilde{b}}_1 \hat{\tilde{b}}_2 \rangle}{dt} =& -i [\Omega_1^{\prime}(t)+\Omega_2^{\prime}(t)] \langle \hat{\tilde{b}}_1 \hat{\tilde{b}}_2 \rangle - \frac{\Gamma_1 + \Gamma_2}{2} \langle \hat{\tilde{b}}_1 \hat{\tilde{b}}_2 \rangle, \\
		\frac{d\langle \hat{\tilde{b}}_1^2 \rangle}{dt} =& [-2 i \Omega_1^{\prime}(t) - \Gamma_1] \langle \hat{\tilde{b}}_1^2 \rangle, \\
		\frac{d\langle \hat{\tilde{b}}_2^2 \rangle}{dt} =& [-2 i \Omega_2^{\prime}(t)-\Gamma_2] \langle \hat{\tilde{b}}_2^2 \rangle.
	\end{align}
\end{subequations}
One can easily find that the values of $\langle \hat{\tilde{b}}_1^\dag \hat{\tilde{b}}_2 \rangle$, $\langle \hat{\tilde{b}}_1 \hat{\tilde{b}}_2 \rangle$, $\langle \hat{\tilde{b}}_1^2 \rangle$, and $\langle \hat{\tilde{b}}_2^2 \rangle$ decay to zero since they are coupling to dissipation but not {subjecting} to any driving. Here, for the {non-zero} terms $\langle \hat{\tilde{b}}_1^\dag \hat{\tilde{b}}_1 \rangle$ and $\langle \hat{\tilde{b}}_2^\dag \hat{\tilde{b}}_2 \rangle$, their dynamics are dominated by $n_{1, \rm th}(t)$ and $n_{2, \rm th}(t)$, respectively, where $n_{1, \rm th}(t)$  $[n_{2, \rm th}(t)]$ is known as a function of the {modified} mechanical frequency  $\Omega_1^\prime(t)$ [$\Omega_2^\prime(t)$].

%%%%%%%%%%%%%%%%%%%%%%
\subsection{Quantum synchronization of dissipative harmonic oscillators}
{Recall the definition of quantum synchronization discussed in Sec.~II, where the complete synchronization of two quantum harmonic oscillators $\hat{b}_1$ and $\hat{b}_2$ is achieved if their standard deviations [$\sigma_{1,x}(t)$, $\sigma_{1,p}(t)$] and [$\sigma_{2,x}(t)$, $\sigma_{2,p}(t)$] satisfy the conditions: $\lim_{t \rightarrow \infty}[\sigma_{1,x}(t)-\sigma_{2,x}(t)]=0$ and $\lim_{t \rightarrow \infty}[\sigma_{1,p}(t)-\sigma_{2,p}(t)]=0$. 
	Here, $\sigma_{j,x}$ and $\sigma_{j,p}$ take the forms}
\begin{equation}
\sigma_{j,x}=\sqrt{\langle \hat{x}_j^2 \rangle-\langle \hat{x}_j \rangle^2}, \, \, \, \, \, \sigma_{j,p}=\sqrt{\langle \hat{p}_j^2 \rangle-\langle \hat{p}_j \rangle^2}, j=1,2.
\end{equation} 
By applying the relations $\hat{x}_j=({\hat{b}_j+\hat{b}^\dag_j})/{\sqrt{2}}$ and $\hat{p}_j=-i({\hat{b}_j-\hat{b}^\dag_j})/{\sqrt{2}}$, and
	separating $\hat{b}_j$ into a classical mean value and a quantum {part}: $\hat{b}_j = \langle \hat{b}_j \rangle +  \hat{\tilde{b}}_j$, the standard deviations $\sigma_{j,x}$ and $\sigma_{j,p}$ of {the} quantum harmonic oscillator $\hat{b}_j$ can be rewritten as
\begin{subequations}\label{sigma}
	\begin{align} 
		\sigma_{j,x}&=\sqrt{\frac{1}{2} + \langle \hat{\tilde{b}}_j^\dag \hat{\tilde{b}}_j\rangle + {\rm Re}[\langle \hat{\tilde{b}}_j^2\rangle]},\\
		\sigma_{j,p} &=\sqrt{\frac{1}{2} + \langle \hat{\tilde{b}}_j^\dag \hat{\tilde{b}}_j\rangle - {\rm Re}[\langle \hat{\tilde{b}}_j^2\rangle]}, j=1,2. 
	\end{align}
\end{subequations}  
It can be seen from Eq.~(\ref{sigma}) that both $\sigma_{j,x}$ and $\sigma_{j,p}$ are functions of $\langle \hat{\tilde{b}}_j^\dag \hat{\tilde{b}}_j \rangle$ and $\langle \hat{\tilde{b}}_j^2\rangle$, thus we have 
	new conditions for quantum complete synchronization 
	\begin{equation}\label{e_condition}
	\lim_{t \rightarrow \infty}{e}_{\rm n_b}(t)=0, \ \ \ \ \  \lim_{t \rightarrow \infty}{e}_{\rm b^2}(t)=0,
	\end{equation}
	where ${e}_{\rm n_b} = \langle \hat{\tilde{b}}_1^\dag \hat{\tilde{b}}_1 \rangle - \langle \hat{\tilde{b}}_2^\dag \hat{\tilde{b}}_2 \rangle$ and ${e}_{\rm b^2} = \langle \hat{\tilde{b}}_1^2\rangle - \langle \hat{\tilde{b}}_2^2\rangle$ stand for the synchronization differences. 
	Also, the equations of motion for $\langle \hat{\tilde{b}}_j^\dag \hat{\tilde{b}}_j \rangle$ and $\langle \hat{\tilde{b}}_j^2\rangle$ can be found in Eq.~(\ref{2nd_eq1}) 
\begin{subequations}\label{dnb} 
	\begin{align}
		{d\langle \hat{\tilde{b}}_j^\dag \hat{\tilde{b}}_j \rangle}/dt &= -\Gamma_j \langle \hat{\tilde{b}}_j^\dag \hat{\tilde{b}}_j \rangle + \Gamma_j n_{j,\rm th}[\Omega^{\prime}_j(t)], \\
		d{\langle \hat{\tilde{b}}_j^2\rangle}/dt &= -[2i\Omega_j^{\prime}(t) + \Gamma_j]\langle \hat{\tilde{b}}_j^2\rangle, j=1,2. 
	\end{align}
\end{subequations}
For complete synchronization, note that all the parameters of two quantum harmonic oscillators are required to be identical: $\Gamma_1=\Gamma_2=\Gamma$, $g_1=g_2=g$, and $\Omega_1=\Omega_2=\Omega$.
Then, one can easily obtain the equations of motion for the synchronization differences from Eq.~(\ref{dnb})
\begin{subequations}\label{denb} 
	\begin{align}
		\dot{e}_{\rm n_b} &= -\Gamma {e}_{\rm n_b}, \\
		\dot{e}_{\rm b^2} &= -[2i\Omega^{\prime}(t) + \Gamma]{e}_{\rm b^2}. 
	\end{align}
\end{subequations}
To check if the quantum synchronization conditions provided by Eq.~(\ref{e_condition}) can be satisfied, we analytically solve Eq.~(\ref{denb}) and obtain the solutions: ${e}_{\rm n_b}(t)= \exp(-\Gamma t){e}_{\rm n_b}(0)$ and 
${e}_{\rm b^2}(t)=\exp{(-2\Gamma t)}\exp{\int^t_0{[-2i\Omega^{\prime}(t')]}dt'}e_{\rm b^2}(0)$. 
We find that both of the synchronization differences ${e}_{\rm n_b}$ and 
${e}_{\rm b^2}$ converge to zero as $t \to \infty$ when $\Gamma>0$. It thus can be concluded that the quantum synchronization led by this approach is asymptotically stable when the quantum harmonic oscillators are subject to dissipation. 

The above discussion has demonstrated that the complete synchronization of two quantum harmonic oscillators can be achieved via dissipative mechanisms induced by the environment. The stability of the complete quantum synchronization induced by the environment has implied that the quantum synchronization can be also realized in chaotic regimes by the same mechanism to be discussed in the next section.

\begin{center}
	\begin{figure}[t]
		\centering
		\includegraphics[width=3.5 in]{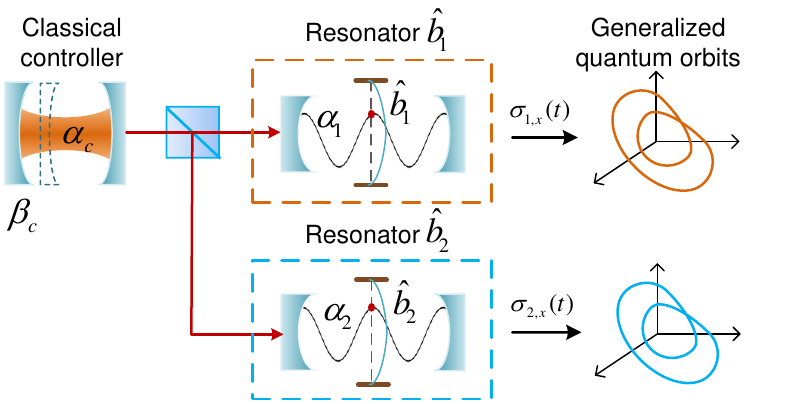}
		\caption{(color online) {An optomechanical setup} for quantum synchronization of two mechanical resonators $\hat{b}_{1}$ and $\hat{b}_{2}$ {in the framework of APD configuration}.}
		\label{Fig3}
	\end{figure}
\end{center}

%%%%%%%%%%%%%%%%%%%%%

\section{A quantum synchronization model: periodic and chaotic motions}
In this section, we study the implementation of the above quantum synchronization model with an  experimentally accessible optomechanical setup. We numerically show that complete synchronization can be achieved in this mode consisting of two quantum mechanical resonators. {Interestingly, this simple model actually allows us to show that the quantum synchronization is robust in either periodic or  chaotic regimes}~\cite{TingYu1999,Naghiloo2017,Mourik2018,RS1,RS2,RS3,random_matrix1,random_matrix2,Haake1991,
	energy_level_1,Neill2016,Slomczynski1994,Zurek1995QuantumCA,Chirikov1995,
	kowalewska2008wigner,Geszti2019,Geszti2018,Cornelius2022,Nakamura1993,Heller2018,semiclassical_1,semiclassical_2,Zurek_1998, Habib1998,Zurek2002,Zurek2003,Bhattacharya2000,Xu2019}.

 As shown in Fig.~\ref{Fig3}, The setup consists of three components: an optomechanical system (left hand) and two quadratic-coupling optomechanical systems (right hand). 
The two quantum mechanical resonators ($\hat{b}_{1}$ and $\hat{b}_{2}$) to be synchronized are distributed in the two separated quadratic-coupling optomechanical systems (right hand), respectively. The left-hand optomechanical system ($\alpha_c$, $\beta_c$) is strongly driven and thus can be treated classically. It outputs a classical field that acts as the inputs of the cavity fields ($\alpha_1$ and $\alpha_2$). The quantum mechanical modes ($\hat{b}_{1}$ and $\hat{b}_{2}$) are modulated by the left-hand optomechanical system ($\alpha_c$, $\beta_c$) via the quadratic coupling with the cavity fields ($\alpha_1$ and $\alpha_2$). {These form an} {\rm APD} configuration that will lead to complete synchronization of {the two quantum mechanical resonators ($\hat{b}_{1}$ and $\hat{b}_{2}$).} Here, the Hamiltonian of the total system reads
\begin{equation}\label{total_hamiltonian}
\begin{split}
H_{\rm total}=&\Delta_c \hat{a}_c^\dag \hat{a}_c + \Omega_c {\hat{b}_c}^\dag \hat{b}_c
+ g_c\hat{a}_c^\dag \hat{a}_c (\hat{b}_c^\dag   + \hat{b}_c) + \varepsilon_c (\hat{a}_c^\dag + \hat{a}_c)\\
&+ \Delta_1 \hat{a}_1^\dag \hat{a}_1 + \Delta_2 \hat{a}_2^\dag \hat{a}_2 + \Omega_2 {\hat{b}_1}^\dag \hat{b}_1
+ \Omega_2 {\hat{b}_2}^\dag \hat{b}_2 \\
&+ g_1\hat{a}_1^\dag \hat{a}_1 \hat{b}_1^\dag \hat{b}_1
+ g_2\hat{a}_2^\dag \hat{a}_2 \hat{b}_2^\dag  \hat{b}_2,
\end{split}
\end{equation}
where $\Delta_k=\omega_k-\omega_{k,d}$ is the detuning between the resonant frequency $\omega_k$ and the external driving $\omega_{k,d}$ of the cavity mode $\hat{a}_k$ for $k=1,2,c$; and its driving strength is denoted by $\varepsilon_k$. Here, the resonant frequency and the damping rate of the mechanical mode $\hat{b}_k$ are $\Omega_k$ and $\Gamma_k$, respectively; while $g_k$ is the optomechanical coupling strength between the cavity mode $\hat{a}_k$ and the mechanical resonator $\hat{b}_k$.

Here, the equations of motion of each cavity (mechanical) mode are described by the Langevin equations
\begin{subequations}
	\label{Langevin1}
	\begin{align}
		\dot{\hat{a}}_c &=-i\Delta_c \hat{a}_c -ig_c \hat{a}_c (\hat{b}_c^\dag + \hat{b}_c) - \frac{\gamma_c}{2} \hat{a}_c+\varepsilon_c
		-\sqrt{\gamma_c}\,\hat{a}_{c,\rm in}, \;
		\\
		\dot{\hat{b}}_c &=-i \Omega_{c}\hat{b}_c-ig_c \hat{a}_c^\dag \hat{a}_c -
		\frac{\Gamma_{\!c}}{2} \hat{b}_c - \sqrt{\Gamma_{\!c}}\, \hat{b}_{c,\rm
			in}, \;
		\\
		\dot{\hat{a}}_1 &= -i \Delta_1 \hat{a}_1 - \frac{\gamma_1}{2}\hat{a}_1 - i g_1 \hat{a}_1 \hat{b}_1^\dag  \hat{b}_1 - \sqrt{\gamma_1 \gamma_c}\,\hat{a}_c-\sqrt{\gamma_1}\hat{a}_{1,\rm in}, \;
		\\
		\dot{\hat{a}}_2 &= -i \Delta_2 \hat{a}_2 - \frac{\gamma_2}{2}\hat{a}_2- ig_2 \hat{a}_2 \hat{b}_2^\dag  \hat{b}_2 - \sqrt{\gamma_2 \gamma_c}\,\hat{a}_c-\sqrt{\gamma_2}\hat{a}_{2,\rm in}, \;
		\\
		\dot{\hat{b}}_1 &= -i \Omega_1 \hat{b}_1 - \frac{\Gamma_1}{2}\hat{b}_1- ig_1 \hat{a}_1^\dag  \hat{a}_1 \hat{b}_1^\dag -\sqrt{\Gamma_1}\hat{b}_{1,\rm in}, \;
		\\
		\dot{\hat{b}}_2 &= -i \Omega_2 \hat{b}_2 - \frac{\Gamma_2}{2}\hat{b}_2- ig_2 \hat{a}_2^\dag  \hat{a}_2 \hat{b}_2^\dag -\sqrt{\Gamma_2}\hat{b}_{2,\rm in}, \;
	\end{align}
\end{subequations}
where $\gamma_k$ and $\hat{a}_{k,\rm in}$ are the damping rate and the input of the optical cavity $\hat{a}_k$, and $\hat{b}_{k,\rm in}$ and  $\Gamma_{\!k}$ are the input and the damping rate of the mechanical mode $\hat{b}_k$ for $k=1,2,c$. 

In this setup, the optomechanical resonator ($\hat{a}_c$, $\hat{b}_c$) and the cavity modes ($\hat{a}_1$ and $\hat{a}_2$) {can be considered classically.}  
By replacing the quantum operators with their classical averages in Eq.~(\ref{Langevin1}): $\alpha_1=\langle \hat{a}_1 \rangle$, $\alpha_2=\langle \hat{a}_2 \rangle$, $\alpha_c=\langle \hat{a}_c \rangle$, and $\beta_c=\langle \hat{b}_c \rangle$, we can then obtain the equations of motions for the classical parts
\begin{subequations}\label{classical_eq}
	\begin{align}
		\dot{\alpha}_c &=-i\Delta_c \alpha_c - \frac{\gamma_c}{2}
		\alpha_c-i g_c \alpha_c (\beta_c + \beta_c^*)
		+\varepsilon_c,
		\\
		\dot{\beta}_c &= \left(-i \Omega_c - \frac{\Gamma_c}{2}\right) \beta_c - i g_c |\alpha_c|^2,
		\\
		\dot{\alpha}_1 &=-i\Delta_1 \alpha_1 - \frac{\gamma_1}{2}
		\alpha_1
		-\sqrt{\gamma_1 \gamma_c}{\alpha}_c + \varepsilon_1,
		\\
		\dot{\alpha}_2 &=-i\Delta_2 \alpha_2 - \frac{\gamma_2}{2}
		\alpha_2
		-\sqrt{\gamma_2 \gamma_c}{\alpha}_c 
		+ \varepsilon_2.
	\end{align}
\end{subequations}
Note that the cavity modes $\alpha_1$ and $\alpha_2$ are not affected by 
the mechanical resonators $\hat{b}_1$ and $\hat{b}_2$. This is because the latter does not contain driving terms, as such the classical averages $\beta_1$ and $\beta_2$ converge to zero as the time $t$ increases. 

Now, we focus on the dynamical evolution of the quantum mechanical resonator $\hat{b}_1$ and $\hat{b}_2$. Since the optomechanical resonator ($\hat{a}_c$, $\hat{b}_c$) and the cavity modes ($\hat{a}_1$ and $\hat{a}_2$) treated classically are omitted in the total Hamiltonian [Eq.~(\ref{total_hamiltonian})], we have the effective system Hamiltonian,
\begin{equation}\label{H_eff_II}
\tilde{H}_{\rm eff}=\hbar\Omega^{\prime}_1(t) \hat{b}_1^\dag \hat{b}_1
+ \hbar\Omega^{\prime}_2(t) \hat{b}_2^\dag \hat{b}_2, 
\end{equation}
where $\Omega^{\prime}_j=\Omega_j + g_j|\alpha_j|^2$ is the modified mechanical frequency due to the optomechanical coupling with the classical optical field $\alpha_j$ for  {$j=1,2$}. 
Here, the cavity mode $\alpha_1$ ($\alpha_2$) links both the classical optomechanical system ($\alpha_c$, $\beta_c$) and the quantum mechanical mode $\hat{b}_1$ ($\hat{b}_2$) together. The chaos generated by the optomechanical system ($\alpha_c$, $\beta_c$), is thus transferred into the quantum mechanical {resonators $\hat{b}_1$ and $\hat{b}_2$}.

{Remind that} the dynamics of the quantum mechanical mode $\hat{b}_{1}$ ($\hat{b}_{2}$), is described by the temporal evolution of its standard deviations ($\sigma_{1,x}$, $\sigma_{1,p}$) [($\sigma_{2,x}$, $\sigma_{2,p}$)], which are  given in Eq.~(\ref{sigma}). 
Specifically, in this quadratic-coupling optomechanical setting, the mean value of the mechanical resonator $\hat{b}_{1}$ ($\hat{b}_{2}$) is always zero: $\langle\hat{b}_1\rangle=0$ ($\langle\hat{b}_2\rangle=0$). Thus, the  standard deviations ($\sigma_{j,x}$, $\sigma_{j,p}$) for the quantum mechanical mode $\hat{b}_{1}$ ($\hat{b}_{2}$) can be rewritten as
\begin{subequations}\label{sigma_2}
	\begin{align} 
		\sigma_{j,x}&=\sqrt{\frac{1}{2} + \langle \hat{b}_j^\dag \hat{b}_j\rangle + {\rm Re}[\langle \hat{b}_j^2\rangle]},\\
		\sigma_{j,p} &=\sqrt{\frac{1}{2} + \langle \hat{b}_j^\dag \hat{b}_j\rangle - {\rm Re}[\langle \hat{b}_j^2\rangle]}, j=1,2, 
	\end{align}
\end{subequations}
where the equations of motions for $\langle \hat{b}_1^\dag \hat{b}_1 \rangle$, $\langle \hat{b}_2^\dag \hat{b}_2 \rangle$, $\langle \hat{b}_1^\dag \hat{b}_2 \rangle$, $\langle \hat{b}_1 \hat{b}_2 \rangle$, $\langle \hat{b}_1^2 \rangle$, and $\langle \hat{b}_2^2 \rangle$  are given by 
\begin{subequations}\label{Linear_eq}
	\begin{align}
		\frac{d\langle \hat{b}_1^\dag \hat{b}_1 \rangle}{dt} =& -\Gamma_1 \langle \hat{b}_1^\dag \hat{b}_1 \rangle+\Gamma_1 n_{1, \rm th}[\Omega_1^\prime(t)], \\
		\frac{d\langle \hat{b}_2^\dag \hat{b}_2 \rangle}{dt} =& -\Gamma_2 \langle \hat{b}_2^\dag \hat{b}_2 \rangle+\Gamma_2 n_{2, \rm th}[\Omega_2^\prime(t)], \\
		\frac{d\langle \hat{b}_1^\dag \hat{b}_2 \rangle}{dt} =& -i [-\Omega_1^{\prime}(t)+\Omega_2^{\prime}(t)] \langle \hat{b}_1^\dag \hat{b}_2 \rangle-\frac{\Gamma_1+\Gamma_2}{2} \langle \hat{b}_1^\dag \hat{b}_2 \rangle, \\
		\frac{d\langle \hat{b}_1 \hat{b}_2 \rangle}{dt} =& -i [\Omega_1^{\prime}(t)+\Omega_2^{\prime}(t)] \langle \hat{b}_1 \hat{b}_2 \rangle - \frac{\Gamma_1 + \Gamma_2}{2} \langle \hat{b}_1 \hat{b}_2 \rangle, \\
		\frac{d\langle \hat{b}_1^2 \rangle}{dt} =& [-2 i \Omega_1^{\prime}(t) - \Gamma_1] \langle \hat{b}_1^2 \rangle, \\
		\frac{d\langle \hat{b}_2^2 \rangle}{dt} =& [-2 i \Omega_2^{\prime}(t)-\Gamma_2] \langle \hat{b}_2^2 \rangle.
	\end{align}
\end{subequations}
One can easily find that the values of $\langle \hat{\tilde{b}}_1^\dag \hat{\tilde{b}}_2 \rangle$, $\langle \hat{\tilde{b}}_1 \hat{\tilde{b}}_2 \rangle$, $\langle \hat{\tilde{b}}_1^2 \rangle$, and $\langle \hat{\tilde{b}}_2^2 \rangle$ decay to zero as the time $t$ increases. For the inhomogeneous equations  $\langle \hat{\tilde{b}}_1^\dag \hat{\tilde{b}}_1 \rangle$ ($\langle \hat{\tilde{b}}_2^\dag \hat{\tilde{b}}_2 \rangle$), the time evolution of these terms will be sensible to $n_{1, \rm th}$ ($n_{2, \rm th}$), where $n_{j, \rm th}(t)=\exp{[{\hbar\Omega_j^\prime}(t)/{\kappa_{\rm B}T}-1]^{-1}}$ is the mean thermal phonon number at the temperature $T$ and the modified resonant frequency $\Omega_j^\prime(t)$ in this setting. 

\begin{center}
	\begin{figure}[t]
		\centering
		\includegraphics[width=3.2 in]{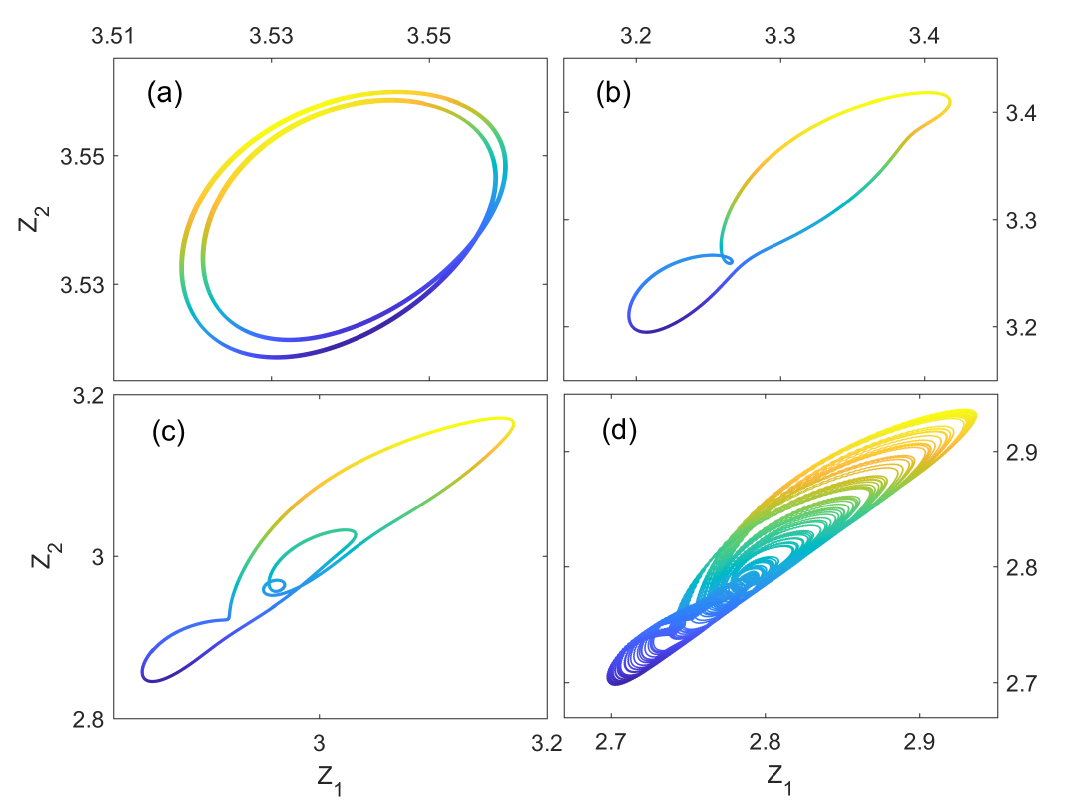}
		\caption{(color online) The phase space orbits of the quantum mechanical resonator $\hat{b}_1$ for different classical optical detunings $\Delta_c/\Omega_c$: (a) $-0.4$, (b) $-0.6$, (c) $-0.85$, and (d) $-0.95$. Other parameters are set as:
			$\gamma_c/\Omega_c=1$, $g_c/\Omega_c=0.001$,
			$\Gamma_c/\Omega_c=0.001$, $\Omega_c/2\pi=1~\rm GHz$, $\varepsilon_c/\Omega_c=418$, $\Delta_1/\Omega_1=-2$, $\gamma_1/\Omega_1=1$, $\varepsilon_1/\Omega_1=0$, $g_1/\Omega_1=0.001$, $\Gamma_1/\Omega_1=10$, $\Omega_1/2\pi=0.01~\rm GHz$, and $T=0.002~\rm k$.}
		\label{Fig4}
	\end{figure}
\end{center}

Also, since $\lim_{t  \rightarrow \infty}{\rm Re}[\langle \hat{b}_j^2\rangle]=0$ ($j=1,2$), we have the relation $\sigma_{j,x}(t)=\sigma_{j,p}(t)$ from Eq.~(\ref{sigma_2}), the condition for quantum synchronization in this setting is simplified as
\begin{equation}\label{condition2}
\lim_{t \rightarrow \infty}[\sigma_{1,x}(t)-\sigma_{2,x}(t)]=0.
\end{equation}
Here, the temporal evolutions of the standard deviations $\sigma_{j,x}$ and $\sigma_{j,p}$ are determined by $\langle \hat{b}_j^\dag \hat{b}_j \rangle$, whose equations of motion are governed by
	\begin{equation}\label{db2} 
			\frac{d\langle \hat{b}_j^\dag \hat{b}_j \rangle}{dt} = -\Gamma_j \langle \hat{b}_j^\dag \hat{b}_j \rangle+\Gamma_j n_{j, \rm th}[\Omega_j^{\prime}(t)], j=1,2.
	\end{equation}
To achieve the complete synchronization, the parameters and the classical inputs of the two quantum mechanical modes $\hat{b}_{1}$ and $\hat{b}_{2}$ are assumed to be identical: $\Omega_1=\Omega_2$, $\Gamma_1=\Gamma_2$, and $\alpha_1(t) \equiv \alpha_2(t)$.  Below we will show numerically that quantum synchronization can be reached in $\hat{b}_{1}$ and $\hat{b}_{2}$ for various settings.

In our simulations, we first prepare the classical controller ($\alpha_c$, $\beta_c$) to four different regimes by adjusting its detuning $\Delta_c/\Omega_c$: 1-period ($-0.4$), 2-period ($-0.6$), 4-period ($-0.85$), and chaos ($-0.95$). For each case, the initial conditions are set to be different:  $\sigma_{1,x}=\sqrt{1.5}$ and $\sigma_{2,x}=\sqrt{10.5}$. The corresponding phase space orbits of the quantum mechanical resonator $\hat{b}_1$ are shown in Fig.~\ref{Fig4}, we find that the quantum dynamics is dominated by the classical controller ($\alpha_c$, $\beta_c$), which was yet presented in Ref.~\cite{Nan2020}. Here, the phase portraits of the quantum system are reconstructed from the time-delayed coordinates $\sigma_{1,x}(\tau), \sigma_{1,x}(2\tau), ..., \sigma_{1,x}(N\tau)$, where $\tau=0.3~{\rm ns}$.

\begin{center}
	\begin{figure}[t]
		\centering
		\includegraphics[width=3.2 in]{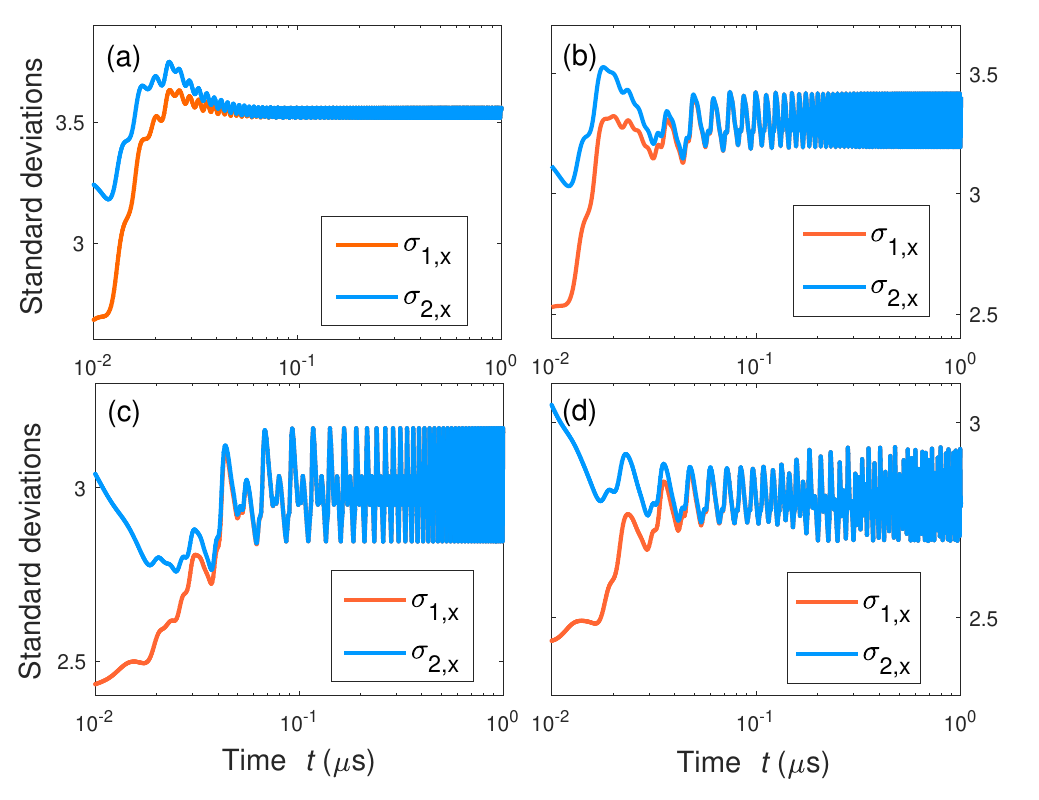}
		\caption{(color online) The standard deviations $\sigma_{1,x}$ and $\sigma_{2,x}$ of the quantum mechanical resonators $\hat{b}_1$ and $\hat{b}_1$ for different quantum motions given in Fig.~\ref{Fig4}. Here, the parameters are set as: $\Delta_1/\Omega_1=\Delta_2/\Omega_2=-2$, $\gamma_1/\Omega_1=\gamma_2/\Omega_2=1$, $\varepsilon_1/\Omega_1=\varepsilon_2/\Omega_2=0$, $g_1/\Omega_1=g_2/\Omega_2=0.001$, $\Gamma_1/\Omega_1=\Gamma_2/\Omega_2=10$, $\Omega_1/2\pi=\Omega_2/2\pi=0.01~\rm GHz$. Other parameters are the same as in Fig.~\ref{Fig4}.} 
		\label{Fig5}
	\end{figure}
\end{center}

%For each quantum regime given in Fig.~4, we numerically calculate the values of $\sigma_{1,x}$ and $\sigma_{2,x}$ of the mechanical mode .
 
The complete synchronization of two quantum mechanical resonators $\hat{b}_{1}$ and $\hat{b}_{2}$ are presented in Fig.~\ref{Fig5}. For each quantum regime given in Fig.~\ref{Fig4}, the values of $\sigma_{1,x}(t)$ and $\sigma_{2,x}(t)$ merge together as the time increases whereas they start from different initial conditions. The complete synchronization is shown to be robust for different quantum motions.
Above numerical results are consistent with the analytic proof given in Sec.~III. The quantum synchronization is shown to be realizable in both stable and unstable regimes, e.g., quantum chaos in Fig.~\ref{Fig5}(d).

\begin{center}
	\begin{figure}[t]
		\centering
		\includegraphics[width=3.4 in]{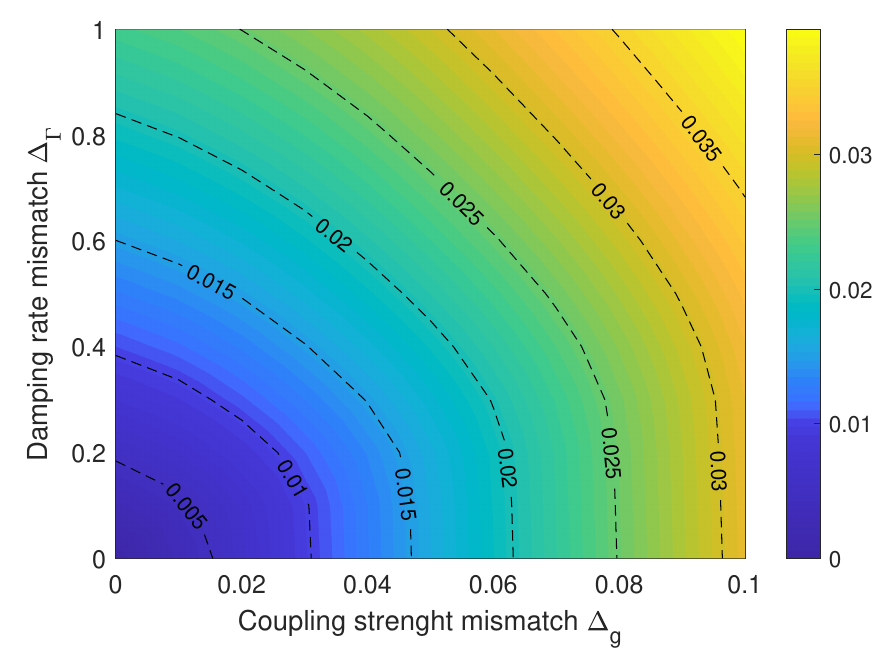}
		\caption{(color online) Average synchronization error $E_{\rm avg}$ for different mismatched parameters $\Delta_{\Gamma}$ and $\Delta_G$. Here, $\Delta_c/\Omega_c=-1$, $g_1=10$ {\rm MHz}, $\Gamma_1=0.15$ {\rm GHz}, and $\Omega_1=\Omega_2=10$ {\rm MHz}. Other parameters are the same as in Fig.~\ref{Fig4}.}
		\label{Fig6}
	\end{figure}
\end{center}

Moreover, we consider the cases that two quantum mechanical modes $\hat{b}_1$ and $\hat{b}_2$ have different parameters. To be more specific, here, we define the mismatched damping rate and mismatched resonant frequency as $\Delta_{\Gamma} = (\Gamma_1-\Gamma_2)/\Gamma_1$ and $\Delta_G = (G_1-G_2)/G_1$. 
Then, we introduce the average synchronization error $E_{\rm avg}$ to measure the effect brought by these mismatched parameters: $E_{\rm avg}=\|\int^{\infty}_{t_0}{\bf e}(t)dt/\int^{\infty}_{t_0}\sigma_{1,x}(t)dt \|$, where $t_0$ is the initial time. 
As shown in Fig.~\ref{Fig6}, the values of $E_{\rm avg}$ are plotted in the $\Delta_{\Gamma}\mbox{-}\Delta_G$ plane and characterized by different colors. It can be seen that $E_{\rm avg}$ is less than $0.01$ even when the mismatched damping rate $\Delta_\Gamma$ is as high as $0.4$; and $E_{\rm avg}$ is $0.03$ when the mismatched coupling strength $\Delta_G$ reaches the value $0.1$. 
Here, it is shown that the complete synchronization of two quantum mechanical modes is still robust for the mismatched parameters $\Delta_\Gamma$ and $\Delta_g$. 

%%%%%%%%%%%%%%%%%%
\section{Conclusions and discussions}

In this paper, we study the synchronization of a continuous-variable system consisting of two quantum harmonic oscillators coupled to dissipative environments. We show that the Active-Passive Decomposition configuration defined in the classical dynamical systems plays a very important role in the quantum regimes where the quantum synchronization can be realized.  For the physical models under consideration, it is proved that the quantum synchronization is asymptotically stable if the quantum systems are subject to dissipation. Moreover, as an example, an experimentally accessible model based on an optomechanical setup is used to illustrate our approach on the  quantum synchronization process defined in this paper. The numerical simulations have clearly indicated that complete synchronization can be achieved and is robust to small parameter mismatches. It is shown that this quantum synchronization approach is not only robust to limit circles, but also chaotic motions. It is desirable to consider the quantum synchronization in different quantum open systems such as dephasing noise, classical noises and colored noises, which will be left for future publications.

\begin{acknowledgments}
This project is partly supported by ART020-Quantum Technologies Project. 
	
\end{acknowledgments}

\bibliography{references}

\end{document}